\documentclass[showpacs,eqsecnum,aps]{revtex4}
\usepackage{amsmath}
\usepackage{amssymb}
\usepackage{amsfonts}
\def\beq{\begin{equation}}
\def\eeq{\end{equation}}
\def\beqa{\begin{eqnarray}}
\def\eeqa{\end{eqnarray}}

\usepackage[dvips]{graphicx,color}

\begin{document}

\title{Equivalent Hamiltonians}

\author{W. N. Polyzou\footnote{This work supported 
in part by the U.S. Department of Energy, under contract DE-FG02-86ER40286}}
\affiliation{
Department of Physics and Astronomy, The University of Iowa, Iowa City, IA
52242}

\vspace{10mm}
\date{\today}

\begin{abstract}
  I give a characterization of the conditions for two Hamiltonians to
  be equivalent, discuss the construction of the operators that relate
  equivalent Hamiltonians, and introduce variational methods that can
  select Hamiltonians with desirable features from the space of
  equivalent Hamiltonians.
\end{abstract}
\vspace{10mm}


\maketitle
\bigskip

\section{Introduction} 

In this paper I study the freedom available to redefine interactions
without changing the scattering and bound-state observables of a
quantum mechanical system.  I consider how this freedom can be used to
formulate interactions that have advantages in various situations.
The Hamiltonians discussed in this paper are equivalent in the
mathematical sense at all energy scales; this a stronger requirement
than the more flexible notion of equivalence used in effective
field theory that only requires equivalence up to some order in the
expansion parameter.  Even with this more restrictive notion of
equivalence there is a very large class of equivalent Hamiltonians.
 
There are a number of formal methods that start from a set of
high-quality two and three-body interactions and construct an
equivalent set of interactions that fit the same bound-state and
scattering data.  These include renormalization group methods and
methods based on specific unitary transformations that block
diagonalize Hamiltonians \cite{Okubo:1954zz}\cite{Suzuki:1980yp}
\cite{Bogner:2003wn} \cite{Nogga:2004ab}
\cite{Bogner:2006pc}\cite{Epelbaum:1999dj}.  The compelling feature of
all of these methods is that the off-diagonal matrix elements that
couple the high- and low-energy parts of the problem are suppressed in
the transformed interactions.  This leads to a low-energy effective
theories that are approximately decoupled from the high-energy part of
the problem.  This has computational advantages in many-body
calculations.  The price paid is that the transformed Hamiltonian has
new many-body forces involving any number of particles.  This is
similar to what is observed using field redefinitions in effective
field theories, although the transformed theories discussed in this
paper are in principle equivalent to the original theories for all
energies.

In this paper I introduce a method that can be used to provide
independent control of the two, three, and many-body interactions.
Much of the work contained in this paper has been discussed in
\cite{Glockle:1990}.  The approach is to start with the general class
of equivalent interactions.  This is then restricted to a subset that
can be treated variationally.  Positive functionals are introduced
that have minimum values for equivalent potentials with selected
properties.  For example it is possible to design functionals that
select models where the dynamics for energies above some given scale
approximately decouples from the dynamics for energies below some
scale, models that have weak three and four-body interactions or models
that emphasize an approximate symmetry.  

\section{Multichannel scattering theory}

In this section I give a brief summary of multichannel scattering
theory that is relevant for this work.

The Hilbert space, ${\cal H}_1$, for a single particle of mass $m$ and
spin $j$ is the space of square integrable functions of particle's
linear momentum and magnetic quantum number
\beq
\langle \mathbf{p},\mu  \vert \psi \rangle = \psi (\mathbf{p},\mu)
\qquad
\langle \psi \vert \psi \rangle = \int d\mathbf{p} 
\sum_{\mu=-j}^j \vert \psi (\mathbf{p},\mu) \vert^2 < \infty .
\label{b.1}
\eeq 
The $N$-particle Hilbert space is the $N$-fold tensor product of
single-particle Hilbert spaces
\beq
{\cal H} := \otimes_{i=1}^N {\cal H}_i .
\label{b.2}
\eeq
The total linear momentum and total 
Galilean mass of the $N$-particle system are the multiplication
operators
\beq
\mathbf{p}:= \sum_{i=1}^N \mathbf{p}_i \qquad M = \sum_{i=1}^N m_i .
\label{b.3}
\eeq
The $N$-body Hamiltonian $H$ has a $N$-body bound state if the 
center-of-mass Hamiltonian, 
\beq
h := H - \frac{\mathbf{p}^2}{2 M}, 
\label{b.4}
\eeq
has a discrete eigenvalue, $-\epsilon$.  If the Hamiltonian is
rotationally and translationally invariant it is possible to find
simultaneous eigenstates of $h$, $\mathbf{p}$, the total $N$-body spin
and the projection of the total $N$-body spin on the 3-axis.  I
denote these eigenstates by $\vert (\epsilon_i ,j_i) , \mathbf{p}, \mu
\rangle$, where the index $i$ labels different bound states
when $h$ has more than one bound state.

To define scattering channels let $a$ denote a partition of the $N$-
particles into $n_a$ disjoint non-empty clusters of $n_{a_i}$
particles.  There is a scattering channel $\alpha$ associated with
the partition $a$ if there is a $n_{a_i}$-body bound state in each of
the $n_a$ clusters of the partition $a$.  Channel states
asymptotically look like a collection of $n_a$ mutually
non-interacting bound clusters.

The direct product of the $n_a$ bound states in the
channel $\alpha_i$ 
\beq
\Phi_{\alpha_i} = \vert (\epsilon_1, j_1), \mathbf{p}_1 ,\mu_1 \rangle \times
\cdots \times \vert (\epsilon_{n_a}, j_{n_a}), \mathbf{p}_{n_a}, \mu_{n_a} \rangle
\label{b.5}
\eeq
defines the mapping $\Phi_{\alpha_i}$, called the channel injection operator,
from the channel Hilbert space, 
${\cal H}_{\alpha_i}$, which 
is tensor product of $n_a$ single 
particle Hilbert spaces, 
\beq
{\cal H}_{\alpha_i} = {\cal H}_1 \otimes \cdots \otimes {\cal H}_{n_a}  
\label{b.6}
\eeq
to the $N$-particle Hilbert space by
\[
\Phi_{\alpha_i} \vert \mathbf{f}_{\alpha_i} \rangle = \int \sum_{\mu_a \cdots \mu_{n_a}}  
\vert (h_1, j_1), \mathbf{p}_1 ,\mu_1 \rangle \times
\cdots \times \vert (h_{n_a}, j_{n_a}), \mathbf{p}_{n_a}, \mu_{n_a} \rangle
\times
\]
\beq
f_1( \mathbf{p}_1, \mu_1) \cdots f_{n_a} (\mathbf{p}_{n_a} , \mu_{n_a}
) \prod d \mathbf{p}_i 
\label{b.7}
\eeq 
where $f_j( \mathbf{p}_j, \mu_j ) $ are wave packets describing the 
momentum and spin distribution of the $j^{th}$ asymptotically bound cluster.

The asymptotic Hilbert space, ${\cal H}_f$, is the direct sum of 
all of the channel Hilbert spaces, {\it including} the one-cluster channels
that correspond to $N$-particle bound states,
\beq
{\cal H}_f = \oplus_i {\cal H}_{\alpha_i},  
\label{b.8}
\eeq
and the multichannel injection operator $\Phi:{\cal H}_f \to {\cal H}$ 
is 
\beq
\Phi \vert \mathbf{f} \rangle := 
\sum_i \Phi_{\alpha_i} \vert \mathbf{f}_{\alpha_i} \rangle
\label{b.9}
\eeq
where
\beq
\vert \mathbf{f} \rangle = \oplus_{\alpha_i} 
\vert \mathbf{f}_{\alpha_i} \rangle .
\label{b.11}
\eeq
For each partition $a$ there may be $0$, $1$ or a finite number of channels.

The unitary time evolution operator $U_{\alpha_i}(t)$ on each 
channel subspace, ${\cal H}_{\alpha_k}$  is
\beq
U_{\alpha_k}(t) = e^{-i \sum_j (\mathbf{p}_j^2/2m_j - \epsilon_j)t}   
\label{b.12}
\eeq
where $\mathbf{p}_j$ is the total momentum of the $j$-th bound cluster 
in the channel $\alpha_k$,  $m_j$ is the total mass of the
$j$-th bound cluster of channel $\alpha_k$ and $-\epsilon_j$ is the 
binding energy of the $j$-th bound cluster of channel 
$\alpha_i$.

The asymptotic time-evolution operator on ${\cal H}_f$ 
is the direct sum of the 
channel time-evolution operators
\beq
U_{f}(t) = \oplus_i U_{\alpha_i}(t) .
\label{b.13}
\eeq
The asymptotic Hamiltonian, $H_f$,  is the infinitesimal generator of 
$U_f(t)$.
Multichannel M{\o}ller wave operators 
\beq
\Omega_{\pm} : {\cal H}_f \to {\cal H}
\label{b.14}
\eeq
are defined by the strong limits
\beq
\Omega_{\pm} = \lim_{t \to \pm \infty} 
U (-t) \Phi U_f (t)  
\label{b.15}
\eeq
where $U(t)= e^{-iHt}$ is the time evolution operator on ${\cal H}$. 
The multichannel scattering operator $S:{\cal H}_f \to {\cal H}_f$,
is defined by 
\beq
S = \Omega^{\dagger}_+ \Omega_- .
\label{b.16}
\eeq

In order to indicate the dependence of the wave operator 
$\Omega_{\pm}$ on 
$H,H_f$ and $\Phi$ I use the notation
\beq
\Omega_{\pm} = \Omega_{\pm} (H, \Phi,H_f) 
\qquad
S (H, \Phi,H_f) = \Omega^{\dagger} _{+} (H, \Phi,H_f) 
\Omega_{-} (H, \Phi,H_f) . 
\label{b.17}
\eeq
I say that the scattering theory is asymptotically complete if the 
wave operators satisfy the following completeness relations:
\beq
I_{\cal H} = \Omega_{+} (H, \Phi,H_f) 
\Omega_{+}^{\dagger}  (H, \Phi,H_f) =
\Omega_{-} (H, \Phi,H_f) \Omega_{-}^{\dagger}  (H, \Phi,H_f)
\label{b.18}
\eeq
and
\beq
I_{{\cal H}_f} = \Omega_{+}^{\dagger}  (H, \Phi,H_f) 
\Omega_{+} (H, \Phi,H_f) =
\Omega_{-}^{\dagger}  (H, \Phi,H_f) \Omega_{-} (H, \Phi,H_f) .
\label{b.19}
\eeq
where $I_{{\cal H}}$ and $I_{{\cal H}_f}$ are the identity operators
on ${\cal H}$ and ${\cal H}_f$ respectively.

The intertwining relations,
\beq
H \Omega_{\pm} (H, \Phi,H_f) = \Omega_{\pm} (H, \Phi,H_f) H_f,
\label{b.20}
\eeq
follow directly from the definition
(\ref{b.15}) and lead to energy conservation,
\beq
[H_f, S]_-=0, 
\label{b.21}
\eeq
in the scattering operator.

In all that follows the wave operators are assumed to exist
and satisfy the completeness relations (\ref{b.18}) and (\ref{b.19}).
The two-Hilbert space formulation of multichannel scattering theory
summarized above is equivalent to the standard formulation of
multichannel scattering.  It has the advantage that the notation
allows all channels to be treated simultaneously.

\section{Cluster expansions}

In this section I introduce combinatorial methods to treat cluster
expansions in this work\cite{rota:1995}\cite{Polyzou:1979wf}\cite{Kowalski:1980cg}.  These provide
an efficient notation for computing the many-body interactions
that appear in different equivalent Hamiltonians.

I begin by introducing a useful notation. 
I let ${\cal P}$ denote the set of partitions of $N$
particles into disjoint, non-empty clusters.  I use lower case Latin
letters, $a$, to denote partitions of $N$ particles, $n_a$ to denote
the number of clusters in the partition $a$, and $n_{a_i}$ to denote
the number of particles in the $i^{th}$ cluster of the partition $a$:
\beq
N=  \sum_{i=1}^{n_a} n_{a_i}. 
\eeq
Thus $a=(125)(37)(64)$ is a three-cluster partition of seven particles,
with one three-particle cluster and two two-particle clusters.

There is a natural partial ordering on the set of partitions of
$N$ particles given by 
\beq
a \subseteq b \quad \mbox{or} \quad b \supseteq a 
\label{d.1}
\eeq
if every particle in the same cluster of $a$ is also in the same
cluster of $b$.  For example $a=(125)(37)(64) \subseteq
b=(125)(3467)$.

I let $a \cup b$ denote the least upper bound of $a$ and $b$ with
respect to this partial ordering and $a \cap b$ denote the greatest
lower bound of $a$ and $b$ with respect to this partial ordering.  I
let $1$ denote the unique 1-cluster partition and $0$ denote the 
unique $N$ cluster partition.  For $a=(125)(37)(64)$
and $b=(125)(367)(4)$ these definitions imply $a\cap b =
(125)(37)(4)(6)$, $a\cup b = (125)(3467)$,
$1= (1234567)$, and $0=(1)(2)(3)(4)(5)(6)(7)$.

Next I introduce the operators that translate clusters.
On each of the single-particle Hilbert-spaces, ${\cal H}_i$, there is a
trivial representation of the three-dimensional Euclidean group
\beq
U_i(\mathbf{x},R) \vert \mathbf{p} , \mu \rangle =
\sum_{\mu'=-j}^j \vert R \mathbf{p} , \mu' \rangle
e^{i R \mathbf{p} \cdot \mathbf{x}}
D^j_{\mu' \mu}(R) 
\label{d.2}
\eeq
where $\mathbf{x}$ are parameters of the space translation subgroup and 
$R$ is a rotation.  The matrix $D^j_{\mu' \mu}(R)$ is the ordinary 
Wigner function.  I define 
\[
U_a (\mathbf{x}_1,R_1, \cdots , \mathbf{x}_{n_a},R_{n_a}):= 
\]
\beq
\otimes_{i_1 \in a_1} U_{i_1} (\mathbf{x}_1,R_1) 
\otimes_{i_2 \in a_2} U_{i_2} (\mathbf{x}_2,R_2) 
\cdots 
\otimes_{i_{n_a} \in a_{n_a} } U_{i_{n_a}} (\mathbf{x}_{n_a} ,R_{n_a}) .    
\label{d.2a}
\eeq
These operators perform independent translations and rotations
on the subsystems of particles in each cluster of the partition $a$.

A bounded operator $A$ on the $N$-particle Hilbert space has a cluster
expansion if it can be expressed as a sum of terms associated with
each partition $a$,
\beq
A = \sum_{a\in {\cal P}} [A]_a,
\label{d.4}
\eeq
where the operators
$[A]_a$ are invariant with respect to independent translations and 
rotations of the cluster of the partition $a$ 
\beq
[ [A]_a , U_a (\mathbf{x}_1,R_1, \cdots , \mathbf{x}_{n_a},R_{n_a})]_-=0,
\label{d.5}
\eeq
and vanish when any pair of particles in the same cluster of the partition
$a$ are asymptotically separated:
\beq
\lim_{\vert \mathbf{x}_i-\mathbf{x}_j \vert\to \infty} 
\Vert [A]_a U_b(\mathbf{x}_1,R_1  \cdots ,\mathbf{x}_{n_b}, R_{n_b}) ) 
\vert \psi \rangle \Vert =0 
\qquad b \nsupseteq a .
\label{d.6}
\eeq
Equations (\ref{d.5} and \ref{d.6}) provide a mathematical characterization
of these two properties.  When $A$ is unbounded I will assume that these
equations hold for a suitable dense set of vectors $\vert \psi \rangle$.  

For $b \supseteq a$,   $U_b (\cdots )$ is a 
subgroup of $U_a (\cdots )$ so   
\[
\lim_{\vert \mathbf{x}_i-\mathbf{x}_j \vert\to \infty} 
\Vert [A]_a U_b(\mathbf{x}_1,,R_1  \cdots ,\mathbf{x}_{n_b}, R_{n_b}) 
\vert \psi \rangle \Vert =
\]
\beq
\lim_{\vert \mathbf{x}_i-\mathbf{x}_j \vert\to \infty} 
\Vert  U_b(\mathbf{x}_1,,R_1  \cdots ,\mathbf{x}_{n_b}, R_{n_b}) [A]_a   
\vert \psi \rangle \Vert =
\Vert [A]_a   
\vert \psi \rangle \Vert \qquad b \nsupseteq a . 
\eeq
It follows from (\ref{d.5}) and (\ref{d.6}) 
that if $A$ has a cluster expansion then
\beq
\lim_{\vert \mathbf{x}_i-\mathbf{x}_j \vert\to \infty} 
\Vert (A - \sum_{b \supseteq a }  [A]_a)  
U_a (\mathbf{x}_1,R_1, \cdots , \mathbf{x}_{n_b},R_{n_b})
\vert \psi \rangle \Vert =0 
\label{d.7}
\eeq
which leads to the definition 
\beq
A_b := \sum_{b \supseteq a }  [A]_a ,
\label{d.8}
\eeq
which is the part of $A$ that is invariant with respect to translations
of the individual clusters of $b$, irrespective of the asymptotic properties:
\beq
[ A_b , U_b (\mathbf{x}_1,R_1, \cdots , \mathbf{x}_{n_b},R_{n_b})]_-=0 .
\label{d.8a}
\eeq
This is the part of $A$ that remains after the clusters of the partition 
$b$ are asymptotically separated.

It is also useful to define
\beq
A^b := A- A_b = \sum_{b \nsupseteq a }  [A]_a
\label{d.9}
\eeq
which is the part of $A$ that asymptotically vanishes when the
different clusters of $b$ are asymptotically separated:
\beq
\lim_{\vert \mathbf{x}_i-\mathbf{x}_j \vert\to \infty} 
\Vert A^b U_b(\mathbf{x}_1,R_1,  \cdots ,\mathbf{x}_{n_b}, R_{n_b})  
\vert \psi \rangle \Vert =0 .
\label{d.9a}
\eeq
The incidence matrix, $\delta_{a \supseteq b}$, called the zeta
function of the partial ordering, $a \supseteq b$, has an inverse,
called the M\"obius function, $\delta_{a \supseteq b}^{-1}$ of the
partial ordering, which also vanishes when $a \nsupseteq b$.
The M\"obius function
can be used to express $[A]_a$ in terms of $A_b$:
\beq
[A]_a = \sum_{b \subseteq a} \delta_{a \supseteq b}^{-1} A_b .
\label{d.10}
\eeq
This inverse is explicitly known \cite{Kowalski:1980cg}:
\beq
\delta_{a \supseteq b}^{-1} = 
\left \{
\begin{array}{ll}
(-1)^{n_a} \prod_{i=1}^{n_a}(-)^{n_{b_i}}(n_{b_i}-1)! & a \supseteq b \\
0 & a \nsupseteq b
\end{array}
\right .
\label{d.10a}
\eeq
where $n_{b_i}$ is the number of clusters of $b$ contained in the 
$i^{th}$ cluster of $a$.

The following identities are consequences of the definitions
\beq
(AB)_a = A_a B_a \qquad (AB)^a = A_aB^a + A^a B_a + A^a B^a
\qquad (A_a)_b = A_{a\cap b}.
\label{d.11}
\eeq

If $A$ has a cluster expansion the connected part of $A$ is the 
part of $A$ that vanishes when any pair 
of particles is separated.  It is  
\beq
[A]_1= A-\sum_{a\not={1}} [A]_a .
\label{d.12}
\eeq

Using properties of the M\"obius and zeta functions gives
the following expression for $[A]_1$:
\beq
[A]_1=\sum_{a\in {\cal P}} \sum_{b\in {\cal P}} 
\delta_{{1} \supseteq a}^{-1} \delta_{a \supseteq b}[A]_b =
\sum_{a \in {\cal P}} \delta_{{1} \supseteq a}^{-1} A_a = 
\delta_{{1} \supseteq {1}}^{-1} A_1 +\sum_{a\not= {1}} 
\delta_{{1} \supseteq a}^{-1} A_a =
\label{d.14}
\eeq
\beq 
A+ \sum_{a\not= {1}} \delta_{{1} \supseteq a}^{-1} A_a
\label{d.15}
\eeq
where I have used the identities 
\beq
\delta_{{1} \supseteq {1}}^{-1}=1 \qquad  A_{1}=A .
\label{d.16}
\eeq
I define the coefficients
\beq
{\cal C}_a:= -\delta_{{1} \supseteq a}^{-1}=(-)^{n_a}(n_a-1)!
\label{d.17}
\eeq
from which it follows that
\beq
A=[A]_{{1}}+ \sum_{a\not= {1}} {\cal C}_a A_a .
\label{d.18}
\eeq
This separates the ``connected'' part $[A]_1$ of $A$ from the disconnected 
part, $\sum_{a\not= {1}} {\cal C}_a A_a$, of $A$. 

\section{Scattering equivalences}

Of interest is a characterization of when two $N$-body Hamiltonians
are physically equivalent.  It is customary in the literature to call
two Hamiltonians equivalent if they are related by a unitary
transformation.  This is really insufficient.  For a simple counter
example consider two different short-ranged repulsive two-body
interactions, $V_1$ and $V_2$.  The spectrum and multiplicity of the
of the two-body Hamiltonians are identical. If the wave operators
satisfy the completeness relations (\ref{b.18}-\ref{b.19}) then the
operator
\beq
A = \Omega_+ (H_1,\Phi, H_0)\Omega_+^{\dagger} (H_2,\Phi, H_0)
\label{c.1}
\eeq
is a unitary operator on ${\cal H}$.  It follows from the intertwining
relations, (\ref{b.20}), that $A$ also satisfies $AH_2= H_1A$; however
any two arbitrary repulsive potentials do not give the same phase
shifts.  So even though $H_1$ and $H_2$ are related by a unitary 
transformation, the scattering observables are unrelated. 
For equivalent Hamiltonians I also need to require that the
$S$ matrix remains unchanged and the description of the free particles
remains unchanged.

The two-Hilbert formulation of scattering is useful in this regard.
What is required in general is the unitary equivalence of the 
Hamiltonians 
\beq
H' = A^{\dagger}  H A 
\qquad
AA^{\dagger} =I 
\label{c.2}
\eeq
and $S$-matrix equivalence
\beq
S (H, \Phi,H_f) = S (H', \Phi',H_f),
\label{c.3}
\eeq
where $H_f$ remains unchanged. Recall from the construction of the 
previous section that the operator $\Phi$ also depends on $H$. 

To determine the requirements of $S$-matrix equivalence on $A$ I
use (\ref{b.17}) in (\ref{c.3}) to obtain 
\beq
\Omega_{+}^{\dagger} (H, \Phi,H_f) \Omega_{-} (H, \Phi,H_f)
=
\Omega_{+}^{\dagger} (H', \Phi',H_f) \Omega_{-} (H', \Phi',H_f).
\label{c.4}
\eeq
Using (\ref{b.16}) in (\ref{c.4}) gives the following candidate for $A$:
\beq
A:= \Omega_{+} (H, \Phi,H_f) \Omega_{+}^{\dagger} (H', \Phi',H_f) =
\Omega_{-} (H, \Phi,H_f)\Omega_{-}^{\dagger}  (H', \Phi',H_f) .
\label{c.5}
\eeq
The intertwining property (\ref{b.20}) gives
\beq
A H' = H A .
\label{c.6}
\eeq
Unitarity of $A$ follows from (\ref{b.18}-\ref{b.19}), which also 
can be used to show
\[
\Omega_{+} (H, \Phi,H_f) = \Omega_{+} (H, \Phi,H_f) I_{{\cal H}_f} =
\]
\[
\Omega_{+} (H, \Phi,H_f) \Omega_{+}^{\dagger} (H', \Phi',H_f)
\Omega_{+} (H', \Phi',H_f) =
\]
\beq
A \Omega_{+} (H', '\Phi,H_f) =
\Omega_{+} (AH'A^{\dagger}, A\Phi',H_f) =
\Omega_{+} (H, A\Phi',H_f) .
\label{c.7}
\eeq
Subtracting the left from the right side of (\ref{c.7}) and using 
the definition of the wave operators
gives the identity
\beq
0 = \Omega_{+} (H, \Phi,H_f) - \Omega_{+} (H, A\Phi',H_f) 
\eeq
which is equivalent to
\beq
0 =  \lim_{t \to \infty} \Vert
U(t) [\Phi - A \Phi' ] U_f (t) \vert \mathbf{f} \rangle \Vert =
\lim_{t \to \infty} \Vert 
[\Phi - A \Phi' ] U_f (t) \vert \mathbf{f} \rangle \Vert .
\label{c.8}
\eeq
Similarly using the second equation (\ref{c.4}) gives the corresponding 
relation with the other time limit
\beq
0 =
\lim_{t \to -\infty} \Vert 
[\Phi - A \Phi' ] U_f (t) \vert \mathbf{f} \rangle \Vert .
\label{c.9}
\eeq
The vanishing of {\it both} time limits is important.  The failure of
$S$-matrix equivalence in the case of the two repulsive potentials is
because the two time limits lead to {\it different} unitary operators,
$A_{+} \not= A_{-}$, satisfying (\ref{c.2}).

The asymptotic conditions, (\ref{c.8}-\ref{c.9}), along with the
definition of the operator $A$ (\ref{c.5}), are consequences of the
identity (\ref{c.3}) of the two scattering operators.

Conversely, if both asymptotic conditions, (\ref{c.8}) and (\ref{c.9}),  
hold for some unitary $A$ then 
\beq
\Omega_{\pm} (H, \Phi,H_f) = \Omega_{\pm} (H, A\Phi',H_f) =
A \Omega_{\pm} (H', \Phi',H_f).
\label{c.10}
\eeq
Because this holds for the same $A$ for both time limits it follows that 
\[
S (H, \Phi,H_f) = 
\Omega_{+}^{\dagger} (H, \Phi,H_f) \Omega_{-} (H, \Phi,H_f)
=
\]
\beq
\Omega_{+}^{\dagger} (H', \Phi',H_f) 
A^{\dagger}A
\Omega_{-} (H', \Phi',H_f)=
S (H', \Phi',H_f) .
\label{c.11}
\eeq
This shows that the asymptotic conditions (\ref{c.8}-\ref{c.9}) are
necessary and sufficient conditions for the invariance of the
$S$-matrix.  This result is the content of a theorem in formal
scattering theory due to Ekstein \cite{Ekstein:1960}.

I also need to determine the relation of $\Phi'$ to $A$ and $H$ in
the context of Ekstein's theorem.  I assume that $A$ has a
well-defined cluster expansion and I define an operator $A_a$ by
turning off the parts of $A$ that vanish when the clusters of the
partition $a$ are asymptotically separated.  It follows that if I
turn off the interactions between particles in different clusters of
the partition $a$ that $H'$ will have the following limiting form
\beq
H'= A^{\dagger} H A \to 
H'_a= A_a^{\dagger} H_a A_a 
\label{c.12}
\eeq
where $H'_a$ is a sum of transformed subsystem Hamiltonians associated
with each cluster.  It follows from the definitions (\ref{b.7}) and
(\ref{b.9}) that the channel injection operators $\Phi_{\alpha_i}$ and
$\Phi'_{\alpha_i}$ associated with the partition $a$ 
are related by
\beq
\Phi'_{\alpha_i} = A^{\dagger}_a \Phi_{\alpha_i}
\label{c.13}
\eeq
where $\Phi'_{\alpha_i}$ is an eigenstate of $H_a'$
with eigenvalues
\beq
E_{\alpha_i} =\sum_{j=1}^{n_a}( \mathbf{p}_j^2/2m_j - \epsilon_j) .
\label{c.14}
\eeq
Thus 
\beq   
\Phi' = \sum_{i} A_{a_i}^{\dagger} \Phi_{\alpha_i} . 
\label{c.14a}
\eeq

Finally, if I want the subsystem Hamiltonians to be separately
rotationally and translationally invariant, then each of the operators
$A_a$, obtained from $A$ by turning off the parts of $A$ that generate
interactions between particles in different clusters of the partition
$a$ should also be translationally and rotationally invariant.

Given a unitary $A$ with a cluster expansion, equations (\ref{c.13}) and
(\ref{c.14a})
imply 
\beq   
A \Phi' = \sum_{i} A A_{a_i}^{\dagger} \Phi_{\alpha_i} . 
\label{c.15}
\eeq
The requirement that $H$ and $H' = A^{\dagger}HA$ give the same $S$ matrix 
is that $A$ is a unitary transformation with a well-defined cluster
expansion satisfying the asymptotic conditions 
\beq
0 =
\lim_{t \to \pm \infty} \Vert 
\sum_i [I - AA_{a}^{\dagger}] \Phi_{\alpha_i}  U_{\alpha_i} (t) \vert 
\mathbf{f}_{\alpha_i} \rangle \Vert =0
\label{c.16}
\eeq
for each channel, 
or equivalently because of the unitarity of $A$
\beq
\lim_{t \to \pm \infty} \Vert 
\sum_i [A^{\dagger}  - A_{a}^{\dagger}] \Phi_{\alpha_i}  U_{\alpha_i} (t) \vert 
\mathbf{f}_{\alpha_i} \rangle \Vert =0 .
\label{c.17}
\eeq
If I assume that all of the $\mathbf{f}_i$ vanish except for the $N$-body 
breakup channel then (\ref{c.16}) implies
\beq
0 =
\lim_{t \to \pm \infty} \Vert 
\sum_i [I - A^{\dagger}]  U_{0} (t) \vert 
\mathbf{f}_{0} \rangle \Vert =0 .
\label{c.18}
\eeq
When $A$ has a suitable cluster expansion, (\ref{c.18}) implies 
(\ref{c.16}).  This is discussed in section V.

Equation (\ref{c.18}) is equivalent to 
\beq
0 =
\lim_{t \to \pm \infty} \Vert 
\sum_i [I - A]  U_{0} (t) \vert 
\mathbf{f}_{0} \rangle \Vert =0 .
\label{c.19}
\eeq
where we have used $\Phi_0=I$ and ${\cal H}_0={\cal H}$ 
for the unique $N$-cluster breakup channel.

I refer to unitary transformations $A$ satisfying (\ref{c.19}) as
scattering equivalences.  It is easy to show that with this definition
the set of scattering equivalences form a group with respect to
operator multiplication.

\section{Asymptotic properties} 

It is now possible to construct a parameterized 
set of  scattering 
equivalences.  Because the scattering 
equivalences $A$ are unitary operators,  it follows that $A$ can 
be expressed as the Cayley transform of a Hermitian operator $\Gamma$
\beq
A = \frac{1 - i \Gamma}{1+ i \Gamma} \qquad \Gamma = \Gamma^{\dagger} .
\label{e.1}
\eeq

In what follows I will assume that the Cayley transform, $\Gamma$,
has a cluster expansion,
\beq
\Gamma = \sum_{a\in {\cal P}}  [\Gamma]_a
\label{e.2}
\eeq
where the $[\Gamma]_a$ are Hermitian,  invariant with respect to
translations and rotations of the clusters of $a$, and vanish when any
of the particles in different clusters of $a$ are asymptotically
separated.  Specifically
\beq
\lim_{\vert \mathbf{x}_i-\mathbf{x}_j \vert\to \infty} 
\Vert [\Gamma]_a U_b(\mathbf{x}_1,R_1  \cdots ,\mathbf{x}_{n_b}, R_{n_b}) 
\vert \psi \rangle \Vert =0 
\qquad b \nsupseteq a .
\label{e.3}
\eeq
A sufficient condition to satisfy all of the cluster conditions 
is that $[\Gamma]_a$ and $T[\Gamma]_a T$, where $T$ is the
$N$-body kinetic energy operator, are both compact after one removes
all of the momentum conserving delta functions.  In what 
follows, rather than formally taking the cluster limit (\ref{e.3}), I use a
switching parameter to turn off the parts of the operators that vanish
in the cluster limit.  Thus, to take the limit where the clusters
of a partition $a$ are separated, I formally write
\beq
\Gamma (\lambda) = \Gamma_a+ \lambda \Gamma^a   \qquad \Gamma(1)=\Gamma
\label{e.4}
\eeq
and take the limit that $\lambda \to 0$.  I call this implementation
of cluster properties algebraic clustering\cite{Coester:1982vt}; it separates the
combinatorial aspects of cluster properties from the analytic aspects. 
 
While the cluster expansions are based on asymptotic properties of
the operators $A$ with respect to translations, the limits of 
interest in this paper are the time limits (\ref{c.16}-\ref{c.19}). 
Although I will not get into the technical details of the cluster 
limits, it is important to understand the relation between the 
cluster limit and the time limit. 

If I consider the time limit in equations (\ref{c.17}), it has the form
\[
\lim_{t \to \pm \infty} \Vert 
\sum_i [A^{\dagger}  - A_{a}^{\dagger}] \Phi_{\alpha_i}  U_{\alpha_i} (t) \vert 
\mathbf{f}_{\alpha_i} \rangle \Vert =
\]
\beq
\lim_{t \to \pm \infty} \Vert 
\sum_i A^{a\dagger} e^{-i\sum (\frac{\mathbf{p}_{a_j}^2}{2m_{a_j}}- \epsilon_j) t}  
\Phi_{\alpha_i}  \vert 
\mathbf{f}_{\alpha_i} \rangle \Vert = 0
\label{e.5}
\eeq
where the kinetic energy that appears in the exponent is the 
sum kinetic energies of each cluster of $a$.  
It looks similar to the cluster limit
\beq
\lim_{\vert \mathbf{x}_j - \mathbf{x}_k \vert \to  \infty} \Vert 
\sum_i A^{a\dagger} e^{i\sum \mathbf{p}_{a_j}\cdot \mathbf{x}_j  }  
\Phi_{\alpha_i}  \vert 
\mathbf{f}_{\alpha_i} \rangle \Vert = 0 .
\label{e.6}
\eeq

To understand the relation between the limits in (\ref{e.5}) and (\ref{e.6}) 
I consider first a single degree of freedom. 
Consider the limit where
the $y$-component of cluster $i$ is being translated.  The time limit 
above is bounded by a sum of terms
of the form
\[
\lim_{\lambda \to \infty} \int_{-\infty}^{+ \infty} 
f(p_{iy}) e^{i p_{iy} \lambda }dp_{iy}  =
\]
\beq
\lim_{\lambda \to \infty} \int_{0}^{ \infty} 
f(p_{iy}) e^{i p_{iy} \lambda }dp_{iy}  +  
\lim_{\lambda \to -\infty} \int_{0}^{ \infty} 
f(- p_{iy}) e^{i p_{iy} \lambda }dp_{iy}.    
\label{e.7}
\eeq
Both terms vanish by the Riemann Lebesgue lemma  if   
$f(p_{iy})$ and $f(-p_{iy})$ are absolutely integrable on 
$[0,\infty]$. 

The corresponding time limit (\ref{c.19}) contains 
a term of the form 
\beq
\lim_{\lambda \to \infty} \int_{0}^{+ \infty} 
f(p_{iy}) e^{i {p^2_{iy} \over 2m_i}  \lambda }dp_{iy} .
\label{e.8}
\eeq
If I let $u={p^2_{iy} \over 2m_i}$ the time limit becomes
\beq
\lim_{\lambda \to \infty} \int_{-\infty}^{ \infty} 
(f(\sqrt{2m_iu})+ f(-\sqrt{2m_iu})) 
e^{iu\lambda }\sqrt{{m_i\over {u}}} du  = 
\lim_{\lambda \to \infty} \int_{0}^{+ \infty} 
g(u) 
e^{iu\lambda } du
\label{e.9}
\eeq
where 
\beq
g(u):= (f(\sqrt{2m_iu})+ f(-\sqrt{2m_iu})) \sqrt{{m_i\over {u}}}
\label{e.10}
\eeq
is an absolutely integrable function of $u$ if $f(p_{iy})$ 
is an absolutely integrable function $p_{iy}$. Using an extension of this
same argument it is possible to show that the two limits
(\ref{e.5}) and (\ref{e.6}) are equivalent,  provided $\Phi_{\alpha_i}$, 
$\mathbf{f}_{\alpha_i}$ and $ A^a$ are all suitably well-behaved
(i.e. so the resulting integrand is absolutely integrable).

This means that the time limit associated with a given channel has a
vanishing limit whenever the space limit associated with the same
channel also vanishes.  Once I eliminate the delta functions, the
compactness condition always ensures that
(\ref{e.5}) and (\ref{e.6}) are satisfied.  
If the functions are smooth the
fall-off is faster.

For channels associated with the partition $a$ the operators
$A^a$ must  vanish for both time limits.   
A sufficient condition for this 
to be satisfied for {\it all} partitions $a$ is that 
\beq
\lim_{t \to \pm \infty} \Vert 
(A - I) \Phi_0 U_0(t)    
\Phi_{0}   \vert 
\mathbf{f}_{0} \rangle \Vert = 0
\label{e.11}
\eeq 
for the $N$ cluster partition $0$.  In this case $\Phi_0=I$,
${\cal H}_0={\cal H}$ 
and this condition becomes
\beq
\lim_{t \to \pm \infty} \Vert 
(A - I) U_0(t)    
\Phi_{0}   \vert 
\mathbf{f}_{0} \rangle \Vert = 0
\label{e.12}
\eeq 
This ensures that $I-A$ is a sum
terms that vanish when all particles are asymptotically separated.  
This is the basis of our claim that (\ref{c.19}) implies (\ref{c.17})
and leads to the characterization (\ref{c.19}) of the asymptotic 
properties of scattering equivalences.

\section{Construction} 

To construct a suitable class of operators
$A$ that can be used in variational calculations,  
consider operators $A$, where the Cayley transform has
a cluster expansion
\beq
A = {1 - i \Gamma \over 1+ i \Gamma} \qquad \Gamma = \Gamma^{\dagger} .
\qquad
\Gamma = \sum_{a\in {\cal P}} [\Gamma]_a
\label{e.1a}
\eeq
where each $[\Gamma]_a$ is a Hermitian operator that commutes with 
$U_a(\mathbf{x}_1,R_1,  \cdots ,\mathbf{x}_{n_a}, R_{n_a})$
and satisfies the asymptotic condition 
\beq
\lim_{\vert \mathbf{x}_i-\mathbf{x}_j \vert\to \infty} 
\Vert [\Gamma]_a U_b(\mathbf{x}_1,,R_1  \cdots ,\mathbf{x}_{n_b}, R_{n_b}) ) 
\vert \psi \rangle \Vert =0 
\qquad b \nsupseteq a .
\label{e.2a}
\eeq

I also assume that after the momentum conserving delta functions are
removed, the remainder is a compact operator with respect to the 
internal variables.  This ensures that (\ref{e.2a}) holds. 
The means that the internal part has an expansion
of the form 
\beq
[\Gamma]_a = I \times \hat{[\Gamma]}_a 
\label{e.9a}
\eeq
where $I$ is associated with the delta functions and compact remainder 
has the canonical form 
\beq
\hat{[\Gamma]}_a = \sum_n \vert \xi_{an} 
\rangle \lambda_{an} \langle \xi_{an} \vert 
\label{e.10a}
\eeq
with $\lambda_{an}=\lambda_{an}^*$, 
$\lim_{n \to \infty} \vert \lambda_{an} \vert \to 0$ and 
$\langle \xi_{am} \vert \xi_{an} \rangle = \delta_{mn}$. 
In a variational framework the coefficients $\lambda_{an}$ and the
orthogonal vectors $\vert \xi_{an} \rangle$ can be chosen to depend on
variational parameters.  

In general the operators $[\hat{\Gamma}]_a$, along with the original
Hamiltonian, are the input to any calculation.  In addition, because
of Ekstein's theorem, the transformation leads to scattering equivalent
Hamiltonian  characterized by a scattering equivalence with a
cluster expansion of the form (\ref{d.4}).  The Cayley transform may be
unbounded, but it will have and algebraic cluster expansion of the
above form.

The cluster expansion of $\Gamma$ can be used to generate the cluster 
expansion of $A$.  Since $\Gamma_a$ and $[\Gamma]_a$ are related by the 
M\"obius and zeta functions, it is possible to 
construct $\Gamma_a$ from the $[\Gamma]_b$'s.  

I have 
\beq
\Gamma_a = \sum_{b} \delta_{a \supseteq b} [\Gamma]_b
\label{e.11a}
\eeq
\beq
A_a = {1 - i \Gamma_a  \over 1+  i\Gamma_a }
\label{e.12a}
\eeq
\beq
[A]_a = \sum \delta^{-1}_{a \supseteq b} A_b .
\label{e.13}
\eeq
The $A_a$'s can be computed recursively on the number of clusters
in the partition, starting with $N-1$ cluster partitions.

The nature of the general construction can be illustrated using 
a three-body example.
In this case 
$\Gamma_{(ij)(k)}= [\Gamma]_{(ij)(k)}$.  For the two cluster partitions,
$a=(ij)(k)$, I first solve the integral equation 
\beq
{1 \over i - \Gamma_{(ij)(k)} } = 
{1 \over i }  - i  [\Gamma]_{(ij)(k)}
{1 \over i - \Gamma_{(ij)(k)} }.     
\label{e.14}
\eeq
For finite rank $[\Gamma]_{(ij)(k)}$ this is an algebraic problem.
$[\Gamma]_a$'s of the form (\ref{e.10a}) can be uniformly approximated by 
finite rank $[\Gamma]_a$.
For the special case that $[\Gamma]_a = \lambda \Pi_{a}$ is a real 
constant multiplied by the direct product of the identity (in
the conserved momentum variables) 
and
a rank-one projection operator,  equation (\ref{e.14})
can be solved analytically.  The solution is
\beq
A_{(ij)(k)} = {1 -i [\Gamma]_{(ij)(k)} \over 1 + i [\Gamma]_{(ij)(k)} } =  
I - { 2 i \lambda \over 1+ i \lambda} \Pi_{(ij)(k)} .
\label{e.15}
\eeq
To use these solutions to compute $A$ I define
\beq
R := {1  \over 1+  i\Gamma  }
\label{e.16}
\eeq
\beq
R_{(ij)(k)} = {1 \over I + i [\Gamma]_{(ij)(k)}} 
\label{e.17}
\eeq
and
\beq
R_{(1)(2)(3)} = I  .
\label{e.18}
\eeq
For the special case of a rank one $\Gamma_{(ij)(k)}$ 
\beq
R_{(ij)(k)} = I - i {\lambda_{(ij)(k)} \over 1 + i \lambda_{(ij)(k)} } 
\Pi_{(ij)(k)}.
\label{e.19}
\eeq

In general the operators $R_{(ij)(k)}$ and $R$ satisfy the resolvent identities
\beq
R_{(ij)(k)} = I  -i [\Gamma]_{(ij)(k)}  R_{(ij)(k)}
\label{e.20}
\eeq
and
\beq
R = R_{(ij)(k)} - i R_{(ij)(k)} ([\Gamma]_{(jk)(i)} + [\Gamma]_{(ki)(j)} +
[\Gamma]_{(123)}
) R 
\label{e.21}
\eeq
\beq
R = R_{(1)(2)(3)} + i \Gamma R 
\label{e.22}
\eeq
To get an equation for $R$ note that (\ref{d.10a}) and (\ref{d.17})  imply
\beq
\sum_{a \not= 1} {\cal C}_a  = 1 .
\label{e.23}
\eeq
Using this with equations (\ref{d.17}) gives the following equation for
$R$: \beq
R = \sum_{a \not= 1} {\cal C}_a  R = \sum_{a \not= 1} {\cal C}_a  R_a  - i  
\sum_{a \not= 1} {\cal C}_a R_a \Gamma^a R .
\label{e.24}
\eeq
Equation (\ref{e.24}) is valid for any number of particles.
For the three-particle case
the driving term and kernel of (\ref{e.24}) can be expressed
in terms of the $[\Gamma]_a$ as
\[
\sum_{a \not= 1} {\cal C}_a  R_a =
R_{(ij)(k)} + R_{(ij)(k)} + R_{(ij)(k)}- 2I =
\]
\beq
I  -i [\Gamma]_{(12)(3)}  R_{(12)(3)}
-i [\Gamma]_{(23)(1)}  R_{(23)(1)}
-i [\Gamma]_{(31)(2)}  R_{(31)(2)}
\label{e.25}
\eeq
and 
\[
- i  
\sum_{a \not= 1} {\cal C}_a R_a \Gamma^a = 
\]
\[
-i (I  -i [\Gamma]_{(12)(3)}  R_{(12)(3)})
([\Gamma]_{(23)(1)} + [\Gamma]_{(31)(2)} + [\Gamma]_{(123)})
\]
\[
-i (I  -i [\Gamma]_{(23)(1)}  R_{(23)(1)})
([\Gamma]_{(31)(2)} + [\Gamma]_{(12)(3)} + [\Gamma]_{(123)})
\]
\[
-i (I  -i [\Gamma]_{(31)(2)}  R_{(31)(2)})
([\Gamma]_{(12)(3)} + [\Gamma]_{(23)(1)} + [\Gamma]_{(123)})
\]
\beq
+ i2I ([\Gamma]_{(12)(3)} + [\Gamma]_{(23)(1)} 
[\Gamma]_{(31)(2)} + [\Gamma]_{(123)})
\label{e.26}
\eeq
\[
=
-i [\Gamma]_{(123)} 
- [\Gamma]_{(12)(3)}  R_{(12)(3)}
([\Gamma]_{(23)(1)} + [\Gamma]_{(31)(2)} + [\Gamma]_{(123)})
\]
\[
-[\Gamma]_{(23)(1)}  R_{(23)(1)}
([\Gamma]_{(31)(2)} + [\Gamma]_{(12)(3)} + [\Gamma]_{(123)})
\]
\beq
-
[\Gamma]_{(31)(2)}  R_{(31)(2)}
([\Gamma]_{(12)(3)} + [\Gamma]_{(23)(1)} + [\Gamma]_{(123)}).
\label{e.27}
\eeq
The important observation is that this operator, which is the kernel of
the integral equation (\ref{e.24}), is compact after delta functions that
arise from overall translational invariance are removed.  It follows that
equation (\ref{e.24}) can be solved my standard Fredholm methods.   
The solution can then be used to construct $A$ using  
\beq
A = {1-i \Gamma \over 1+ i \Gamma} = 
(1-i \Gamma ) R =
\label{e.28}
\eeq
\beq
(I -i ([\Gamma]_{(12)(3)} + [\Gamma]_{(23)(1)} +
[\Gamma]_{(31)(2)} +  [\Gamma]_{(123)}) R .
\label{e.29}
\eeq

If the individual $[\Gamma]_{a}$ are finite rank (after all of the delta
functions due to the translational symmetry are removed) then it
follows that the kernel (\ref{e.27}) is finite rank (after the
overall momentum conserving delta function is removed) . This is most
easily seen in the special case where all of the $[\Gamma]_a$ are
proportional to one-dimensional projectors (after the delta functions
are removed).  In this case the kernel (\ref{e.27}) becomes
\[
-i \lambda_{(123)}\Pi_{(123)} 
- {\lambda_{(12)(3)} \over 1 + i\lambda_{(12)(3)}} \Pi_{(12)(3)})
(\lambda_{(23)(1)}\Pi_{(23)(1)} + \lambda_{(31)(2)}\Pi_{(31)(2)}
\lambda_{(123)}\Pi_{(123)}
\]
\[
- {\lambda_{(23)(1)} \over 1 + i\lambda_{(23)(1)}} \Pi_{(23)(1)})
(\lambda_{(31)(2)}\Pi_{(31)(2)} + \lambda_{(12)(3)}\Pi_{(12)(3)}
\lambda_{(123)}\Pi_{(123)}
\]
\beq
- {\lambda_{(31)(2)} \over 1 + i\lambda_{(31)(2)}} \Pi_{(31)(2)})
(\lambda_{(12)(3)}\Pi_{(12)(3)} + \lambda_{(23)(1)}\Pi_{(23)(1)}
\lambda_{(123)}\Pi_{(123)} .
\label{e.30}
\eeq
This is a finite dimensional matrix involving the ten operators
$\Pi_{(123)}$,
$\Pi_{(12)(3)} \Pi_{(23)(1)}$, 
$\Pi_{(12)(3)} \Pi_{(31)(2)}$, 
$\Pi_{(12)(3)} \Pi_{(123)}$, 
$\Pi_{(23)(1)} \Pi_{(31)(2)}$, 
$\Pi_{(23)(1)} \Pi_{(12)(3)}$, 
$\Pi_{(23)(1)} \Pi_{(123)}$, 
$\Pi_{(31)(2)} \Pi_{(23)(1)}$, 
$\Pi_{(31)(2)} \Pi_{(12)(3)}$,
and 
$\Pi_{(31)(2)} \Pi_{(123)}$.
After the overall momentum conserving delta function is removed the range of 
the this operator is a ten dimensional vector space.
The resulting integral equation involves solving a system of 10 
linear equations.  When the operators $[\Gamma]_a$ are finite rank,
rather than rank one,  
the matrix is larger, but it is still finite dimensional.

This construction can be extended to any number of particles.  
The kernel of the integral equation for the $N$-body $R$ 
is still finite rank if all of the input $[\Gamma]_a$ are finite rank.
Thus, for finite rank $[\Gamma]_a$ the construction of $A$ involves only 
quadratures and  linear algebra.

Returning to the three-body example have
\beq
H= T + V_{(12)(3)} + V_{(23)(1)} + V_{(31)(2)} + V_{(123)}  
\eeq
\beq
A_{(1)(2)(3)}= I 
\label{e.31}
\eeq
\beq
[A]_{(12)(3)}=  (I - i [\Gamma]_{(12)(3)}) R_{(12)(3)} -I =
-2 i [\Gamma]_{(12)(3)} R_{(12)(3)}
\label{e.32}
\eeq
\beq
[A]_{(23)(1)}=  (I - i [\Gamma]_{(23)(1)}) R_{(23)(1)} -I =
-2 i [\Gamma]_{(23)(1)} R_{(23)(1)}
\label{e.33}
\eeq
\beq
[A]_{(31)(2)}=  (I - i [\Gamma]_{(31)(2)}) R_{(31)(2)} -I =
-2 i [\Gamma]_{(31)(2)} R_{(31)(2)}
\label{e.34}
\eeq
\beq
[A]_{(123)} = 
- i  
\sum_{a \not= 1} {\cal C}_a R_a \Gamma^a R =
\label{e.35}
\eeq
\[
R_{(12)(3)}([\Gamma]_{(23)(1)}+ [\Gamma]_{(31)(2)} + [\Gamma]_{(123)})R +
\]
\[
R_{(23)(1)}([\Gamma]_{(31)(2)}+ [\Gamma]_{(12)(3)} + [\Gamma]_{(123)})R +
\]
\[
R_{(31)(2)}([\Gamma]_{(12)(3)}+ [\Gamma]_{(23)(1)} + [\Gamma]_{(123)})R
\]
\beq
- 2 ([\Gamma]_{(12)(3)}+ [\Gamma]_{(23)(1)} + [\Gamma]_{(31)(2)} +
[\Gamma]_{(123)}) R
\eeq

I can use these cluster expansions of the scattering equivalences
to determine the cluster expansion of the transformed three-body 
Hamiltonian
\beq
H'= A^{\dagger} H A 
\label{e.36}
\eeq
\beq
H_a' = A^{\dagger}_a H_a A_a
\label{e.37}
\eeq
\beq
[H']_{1} = H' - \sum_{a\not=1} {\cal C}_a H_a' = 
A^{\dagger} H A - \sum_{a\not=1} {\cal C}_a A^{\dagger}_a H_a A_a .
\label{e.38}
\eeq
This means that the transformed two-body interactions are
\beq
V'_{(ij)(k)} = [H']_{(ij)(k)}=
(I + [A]^{\dagger}_{(ij)(j)})(T + V_{(ij)(k)} )(I + [A]_{(ij)(j)}) - T  =
\label{e.39}
\eeq
\[
V_{(ij)(k)} + 
[A]^{\dagger}_{(ij)(j)}V_{(ij)(k)} +
[A]^{\dagger}_{(ij)(j)}T +
[A]^{\dagger}_{(ij)(j)}T [A]_{(ij)(j)} +
\]
\beq
[A]^{\dagger}_{(ij)(j)})V_{(ij)(k)}[A]_{(ij)(j)} +
V_{(ij)(k)} [A]_{(ij)(j)} +
T [A]_{(ij)(j)}
\label{e.40}
\eeq
where $T$ is the three-body kinetic energy and the 
$A_{(ij)(k)}$ are given by (\ref{e.32}-\ref{e.34}).

An important observation is that $V'_{(ij)(k)}$ only depends on $T$,
$V_{(ij)(k)}$, and $[\Gamma]_{(ij)(k)}$.  It does not depend on
$[\Gamma]_{(123)}$.  This means that after one chooses
$[\Gamma]_{(ij)(k)}$ to give a transformed two-body interaction, it is
still possible to use the freedom to {\it independently} choose
$[\Gamma]_{(123)}$ to transform the resulting three-body interaction
without changing the transformed two-body interactions.

The transformed three-body interaction is
\beq
V_{(123)'}' = A^{\dagger} H A - T -  
V'_{(12)(3)}- V'_{(23)(1)}- V'_{(31)(2)} =
[A^{\dagger} H A]_1 =
\label{e.41}
\eeq
\[
[A]_{(123)} H A^{\dagger} + 
\]
\[
[A]_{(12)(3)} \left ( (T+ V_{(12)(3)})( [A]^{\dagger}_{(23)(1)}+ 
[A]^{\dagger}_{(31)(2)} + [A]^{\dagger}_{(123)}) + 
(V_{(23)(1)} + V_{(31)(2)} + V_{(123)}) A^{\dagger} \right )+ 
\]
\[
[A]_{(23)(1)} \left ( (T+ V_{(23)(1)})( [A]^{\dagger}_{(31)(2)}+ 
[A]^{\dagger}_{(12)(3)} + [A]^{\dagger}_{(123)}) + 
(V_{(31)(2)} + V_{(12)(3)} + V_{(123)}) A^{\dagger} \right )+ 
\]
\[
[A]_{(31)(2)} \left ( (T+ V_{(31)(2)})( [A]^{\dagger}_{(12)(3)}+ 
[A]^{\dagger}_{(23)(1)} + [A]^{\dagger}_{(123)}) + 
(V_{(12)(3)} + V_{(23)(1)} + V_{(123)}) A^{\dagger} \right )+ 
\]
\[
T[A]^{\dagger}_{(123)} + 
V_{(12)(3)} ([A]^{\dagger}_{(23)(1)} + [A]^{\dagger}_{(31)(2)} + 
[A]^{\dagger}_{(123)}) +
\]
\[
V_{(23)(1)} ([A]^{\dagger}_{(31)(2)} + [A]^{\dagger}_{(12)(3)} + 
[A]^{\dagger}_{(123)}) +
\]
\beq
V_{(31)(2)} ([A]^{\dagger}_{(12)(3)} + [A]^{\dagger}_{(23)(1)} + 
[A]^{\dagger}_{(123)}) .
\eeq
This is expressed as a sum of completely connected terms;
it could be expressed in a more symmetric form but that would involve 
more terms.  The entire expression depends on the operators
$[\Gamma]_a$ that depend on the variational parameters.  

If the $[A]_{(1j)(k)}$ have already been determined by fixing the 
two-body interaction then one can start with the transformed potential and
use an $A$ where only $[\Gamma]_{(123)]}$ is non-zero to get an 
optimized three-body interaction.  Alternatively one could state with 
the original potential and leave the $[A]_{(ij)(k)}$ fixed in the above 
expression, with all of the variational parameters in $[\Gamma]_{(123)}$.

\section{controlling the Hamiltonian}
 
I order to use variational methods to determine the best 
choice of Hamiltonian a positive functional is needed that 
can be minimized.  It is possible to either work recursively
on the number of particles, by first determining two-body 
interactions, followed by the three-body interaction, or 
alternatively to determine all interactions simultaneously. 

The simplest type of functionals are of the form
\beq
F(V) = \mbox{Tr} ( \rho V^{\dagger} V)^{1/2}
\label{f.1}
\eeq
where $\rho$ is a positive, rotationally and translationally invariant 
operator. 
The trace is only taken over the variables that remain {\it 
after} the 
momentum conserving delta functions are removed.  Thus for two-body 
interactions of the form 
\beq
V(\mathbf{k},\mathbf{k}', \eta_1 \cdots \eta_N) \delta (\mathbf{p}'-
\mathbf{p})
\eeq
$F(V)$ would have the general form
\beq
F(V) = \int d\mathbf{k} d\mathbf{k}'d\mathbf{k}'' 
V(\mathbf{k},\mathbf{k}', \eta_1 \cdots \eta_N)
V^*(\mathbf{k}'',\mathbf{k}', \eta_1 \cdots \eta_N)
\rho (\mathbf{k},\mathbf{k}'') 
\eeq
with obvious generalizations for three-body interaction.

\beq
F(V) = \int d\mathbf{k} d\mathbf{k}'d\mathbf{k}'' 
V(\mathbf{k}-\mathbf{k}', \eta_1 \cdots \eta_N)
V^*(\mathbf{k}'-\mathbf{k}'', \eta_1 \cdots \eta_N)
\rho (\mathbf{k}''-\mathbf{k}') 
\eeq

If the starting potential is local this expression has to be modified
because $V V^{\dagger}$ is a function of the difference
$\mathbf{k}'-\mathbf{k}$ which leads to an infinite volume factor.
While local potentials 
can be treated by using a different positive functional,  an alternative is
to note that if $V=V_{loc} + V_r$ then $F(V) = F(V_{loc}) + F'(V_r,V_{loc})$.
It is only the first term that is infinite, but this term does not
depend on the variational parameters.  The second term will be finite
for suitable $A$ and it contains all of the dependence on the variational 
parameters.  It follows that the critical value of the 
variational parameters can be determined by requiring that 
all partial derivatives of the second term at the critical 
value of the parameters.

The general procedure is to start from a given $N$-body Hamiltonian,
$H$, and a parameterized set of scattering equivalences 
$A(\eta_1, \cdots ,\eta_n)$ where $\eta_i$ are variational parameters.
The scattering equivalences $A(\eta_1, \cdots ,\eta_n)$ generate a 
parameterized set 
of equivalent Hamiltonians:
\beq
H'(\eta_1, \cdots ,\eta_n) = A^{\dagger}(\eta_1, \cdots ,\eta_n)
H A (\eta_1, \cdots ,\eta_n).   
\eeq
They have cluster expansions 
\beq
H' = T + \sum_{ij} V'_{ij}(\eta_1, \cdots ,\eta_n)+
\sum_{ijk} V'_{ijk}(\eta_1, \cdots ,\eta_n) + \cdots + 
V'_{N}(\eta_1, \cdots ,\eta_n).
\eeq
The two, three, four $\cdots$ N-body interactions all depend on the 
choice of variational parameters.

For example, to construct two-body interactions that have primarily 
low-momentum content I would choose a functional that is large 
when the momenta are large.  The functional has to be chosen 
so the trace is finite for all interactions in the model space. 

A functional of the form 
\beq
{\rho}( \mathbf{k} , \mathbf{k}')  = 
\mbox{tanh} (\alpha + \mathbf{k}^2/\mathbf{k}_0^2)
\mbox{tanh} (\alpha + \mathbf{k}^{\prime 2}/\mathbf{k}_0^2), 
\label{tanh}
\eeq
where $\alpha$ is a small dimensionless quantity, would suppress
momentum components above the scale $\mathbf{k}_0^2$.
Alternatively I can
design positive functionals that weaken
three-body forces or reduce two-body correlations.

Finding minimum of the functional 
\beq 
\mbox{Tr}( {\rho} V^{\dagger}_{12}(\eta_1, \cdots ,\eta_n) 
V_{12}(\eta_1, \cdots ,\eta_n) )
\eeq
with respect to the parameters $\eta_1, \cdots ,\eta_n$ selects
equivalent potentials that have low-momentum content.  

After the two, three, $\cdots N-$body interactions have been
determined, then I can use the new Hamiltonian as the starting
point.  I can construct a new set of interactions using scattering
equivalences with $[\Gamma]_{(ij)(k)}=0$.  These scattering
equivalences only affect the three and more-body interactions.  I can
choose a new three-body $\rho$ that emphasizes some desirable feature
of the three-body interaction.  The local minimum generates a new
three-body interaction.  Combining the two scattering equivalences
leads to an scattering equivalence $A$ that transforms $H'' =
A^{\dagger} HA$.  

If this is embedded in the $N$-particle Hilbert space it 
(1) generates the selected two and three-body interactions,
(2) new $4,5 \cdots N$-body interactions, and (3) explicit unitary 
transformations, $A$,  that can be used to generate transformed 
operators like electromagnetic current operators
\beq
J^{\mu'}(x) = A^{\dagger} J^{\mu} (x) A .
\eeq

\section{Simple example}

To illustrate the method I consider a two-body Hamiltonian of the form
\beq
H = {\mathbf{k}^2 \over 2 \mu} + V,
\label{g.1}
\eeq
where I assume that $V$ is a local potential.  I consider a 
parameterized rank one unitary transformation of the form 
\beq
A(\lambda)  = I + \vert g \rangle {2 i \lambda \over 1 -i \lambda \langle g \vert g \rangle}\langle g \vert =
I + \vert g \rangle f(\lambda) \langle g \vert
\label{g.2}
\eeq
where $\vert g \rangle$ is a fixed form factor 
and $\lambda$ is a variational 
parameter.  The transformed potential is
\beq
V'(\lambda ) = A^{\dagger}(\lambda)  H A (\lambda) - 
{\mathbf{k}^2 \over 2 \mu} .
\label{g.3}
\eeq
The transformed potential differs from the original potential by the addition 
of a finite number of separable terms.  It has the form 
\beq
V'(\lambda)  = V + \vert g \rangle f^*(\lambda) \langle g \vert H + 
H \vert g \rangle f(\lambda) \langle g \vert
+ \vert g \rangle f^*(\lambda ) \langle g \vert H \vert g \rangle 
f(\lambda) 
\langle g \vert = V + V_r(\lambda) 
\label{g.4}
\eeq
The first term in this expression is local but independent of 
$\lambda$.  The remaining terms are separable and depend on $\lambda$.

I use the density (\ref{tanh}), with a chosen value of $\mathbf{k}_0$. 
It  has the form 
\beq
\rho = \vert \chi \rangle \langle \chi \vert 
\label{g.5}
\eeq
leads to the variational function
\beq
F(\lambda ) := \langle \chi \vert \left ( V^{\dagger\prime}(\lambda) V'(\lambda )
- V^{\dagger} V \right )  
\vert \chi \rangle =
\langle \chi \vert \left ( 
V^{\dagger}V_r(\lambda)+ V_r^{\dagger} (\lambda)V +
V_r^{\dagger} (\lambda)V_r (\lambda) 
\right )  
\vert \chi \rangle .
\label{g.6}
\eeq
The subtracted  contribution, $V^{\dagger}V$, eliminates the infinite
constant that appears for local $V$.  The terms in the resulting expression 
are
\[
F(\lambda ) = 
\]
\[
\langle \chi \vert
V^{\dagger}\vert g \rangle f^*(\lambda) \langle g \vert H \vert \chi \rangle  + 
\langle \chi \vert V^{\dagger} H \vert g \rangle f(\lambda) 
\langle g \vert \chi \rangle
+ 
\langle \chi \vert V^{\dagger} \vert g \rangle f^*(\lambda ) 
\langle g \vert H \vert g \rangle 
f(\lambda) 
\langle g \vert \chi \rangle +
\]
\[
\langle \chi \vert g \rangle f^*(\lambda) 
\langle g \vert H V \vert \chi \rangle + 
\langle \chi \vert H \vert g \rangle 
f(\lambda) \langle g \vert V \vert \chi \rangle + 
\langle \chi \vert g \rangle f^*(\lambda ) \langle g \vert H \vert g \rangle 
f(\lambda) 
\langle g \vert V \vert \chi \rangle + 
\]
\[
\left (\langle \chi \vert g \rangle f^*(\lambda) \langle g \vert H + 
\langle \chi \vert H \vert g \rangle f(\lambda) \langle g \vert
+ \langle \chi \vert  g \rangle f^*(\lambda ) \langle g \vert H \vert g \rangle 
f(\lambda) 
\langle g \vert \right ) \times 
\]
\beq
\left (
\vert g \rangle f^*(\lambda) \langle g \vert H \vert \chi \rangle+ 
H \vert g \rangle f(\lambda) \langle g \vert \chi \rangle
+ \vert g \rangle f^*(\lambda ) \langle g \vert H \vert g \rangle 
f(\lambda) 
\langle g \vert \chi \rangle \right ).
\label{g.7}
\eeq
This has the form
\beq
F(\lambda ) = c_1 f(\lambda) + c_1^* f^*(\lambda ) +
c_2 f(\lambda)f^* (\lambda) + c_3 f^2(\lambda )f^*(\lambda ) +
c_3^* f^{*2}(\lambda )f (\lambda ) +
c_4  (f(\lambda)f^* (\lambda))^2 
\label{g.7.a}
\eeq
with 
\beq
f(\lambda ) = {2i\lambda \over 1-i \lambda \langle g \vert g\rangle} .
\label{g.7.b}
\eeq
The coefficients $c_k$ are liner combinations of 
the integrals 
$\langle \chi \vert
V^{\dagger}\vert g \rangle$,
$\langle g \vert H \vert \chi \rangle$,
$\langle \chi \vert V^{\dagger}H \vert g \rangle$,
$\langle g \vert H V\vert \chi \rangle$,
$\langle g \vert \chi \rangle$,
$\langle g \vert H \vert g \rangle$, 
$\langle g \vert g \rangle $ and 
$\langle g \vert H^2 \vert g \rangle$.  Since these do not involve $\lambda$
they only have to be computed once.  Although $f(\lambda)$ is complex, 
$F(\lambda)$ is a real function of $\lambda$.  The $\lambda$ dependence is a
rational function.

The critical value of $\lambda=\lambda_c$ is determined by solving  
${dF \over d \lambda} (\lambda_c) =0$ for $\lambda_c$.
The resulting transformed Hamiltonian 
\beq
H' = {\mathbf{k}^2 \over 2 \mu} + V'(\lambda_c)  
\label{g.8}
\eeq
gives the same binding energies and phase shifts as the original
potential of any value of $\lambda$.  The critical value of $\lambda$
will lead to a potential that suppress momenta above $\mathbf{k}_0^2$.
Obviously a softer potential will result if a larger class of unitary
transformations $A$ are used.
  
The original Hamiltonian did not have to be diagonalized to find
the new potential.  In this case, by varying $\lambda$ from $0$ to its
critical value it is possible to continuously evolve the initial 
local potential to the final soft potential. 

Since the unitary scattering equivalence is given as an 
explicit operator valued function of $\lambda$, I can calculate how 
observables evolve with the parameter $\lambda$.  For example the  
electromagnetic current operators transforms as follows:
\beq
J^{\mu}(x)' = J^{\mu}(x) + 
f^*(\lambda) \vert g \rangle \langle g \vert J^{\mu}(x) + 
f(\lambda) J^{\mu}(x) \vert g \rangle \langle g \vert +
f^*(\lambda) f(\lambda) \vert g \rangle \langle g \vert J^{\mu}(x)
\vert g \rangle \langle g \vert
\label{g.9}
\eeq
Finally, give the two-body unitary transformation for each pair of
particles, $A_{ij}(\lambda)$, it is 
possible to construct the corresponding three-body unitary operator
following the method of the previous section.  In terms of the 
above parameters, for three identical particles $A$ has the form  
\beq
A = {I - i \alpha \over I + i \alpha} 
\label{g.10}
\eeq
with 
\beq
\alpha = i {f(\lambda) \over 2 + f(\lambda) \langle g \vert g \rangle}
\left ( \vert g_{12} \rangle \langle g_{12} \vert + 
\vert g_{23} \rangle \langle g_{23} \vert + 
\vert g_{31} \rangle \langle g_{31} \vert \right ) .   
\label{g.11}
\eeq
If the symmetric product of this unitary transformation for each 
pair is applied to the corresponding three-body Hamiltonian the 
transformed three-body Hamiltonian will have the form 
\beq
H' = A^{\dagger}(\lambda)H A = 
K+ V_{12}'(\lambda)+ V_{23}'(\lambda) + V_{31}'(\lambda)
+ V_{123}'(\lambda).    
\label{g.12}
\eeq
The three-body force terms will appear even if the original Hamiltonian 
has only two-body forces.   The computation of $A$ from (\ref{g.10})
involves quadratures and linear algebra, as discussed in section 
VI.

The evolution of the current and the three-body Hamiltonian from their 
original to their final values can be determined by varying 
$\lambda$ from zero to the critical value, $\lambda_0$. 

\section{Conclusion}

In this paper I determined conditions that are necessary and
sufficient for two Hamiltonains to be physically equivalent.  I used
the characterization of these unitary operators to construct a large
class of equivalent $N$-body Hamiltonians that depend on variational
parameters.  There is considerable freedom in choosing the space of
equivalent Hamiltonians.  By choosing functions whose local minima
select Hamiltonians with desirable properties from the space of
equivalent Hamiltonians, it is possible to select classes of
equivalent potentials with desirable properties.  The general freedom
available allows for the possibility of selecting two-body
interactions with desirable properties, then subsequently selecting
among equivalent three-body interactions with desirable properties.
This procedure can be continued for any number of particles, allowing 
independent control of the two, three, four, $\cdots$ interactions.
Because the $k$-body parts of $A$ affect all operators with $k$ or more
particles,  one hopes that desirable properties of the $k$-body 
interaction might persist for the $k+m$ body problems.

While in general it is possible to systematically weaken three and more-body 
interactions using these methods, it is not generally possible to 
eliminate them.   The extremal interactions that are generated are
not fundamental, they depend specifically on the choice of 
positive functional that is used to select these interactions.

The selection of equivalent potentials does not require diagonalizing
any Hamiltonians; it only requires finding local minima of some user
defined functionals.  The functionals are designed so they get large
for interactions with undesirable features.  Once the operators
$[\Gamma]_a$ are determined variationally, it is then possible to
construct scattering equivalences $A$ that operate on systems of any
number of particles, and can be used to construct equivalent observables
in the transformed representation.  For a large class of variational
Hamiltonians the operators $A$ can be constructed from the 
$\Gamma_a$ by finite linear algebra.
 
The general method can be combined with other methods, such are
renormalization group methods, to reduce the strength of the
transformed three-body force without changing the transformed two-body
interactions.

The characterization of the group of scattering equivalences
demonstrates the large class of equivalent Hamiltonains that can be
selected by considering only spectral properties and scattering
observables.  This leads to a lot of flexibility in building
equivalent models of the quantum $N$-body problem.

The author would like to acknowledge support for this work by the U.~S.
Department of Energy, Office of Nuclear Physics, under contract
No. DE-FG02-86ER40286.


\end{document}